 \documentclass[preprint]{aastex}

\shortauthors{Han et al.}
\shorttitle{The Presence of Two Distinct RGBs in NGC~1851}

\begin{document}

\title{THE PRESENCE OF TWO DISTINCT RED GIANT BRANCHES IN THE GLOBULAR CLUSTER NGC~1851}

\author{
Sang-Il Han\altaffilmark{1},
Young-Wook Lee\altaffilmark{1,4},
Seok-Joo Joo\altaffilmark{1},
Sangmo Tony Sohn\altaffilmark{2},
Suk-Jin Yoon\altaffilmark{1},
Hak-Sub Kim\altaffilmark{1}, and
Jae-Woo Lee\altaffilmark{3}
}

\altaffiltext{1}{Center for Space Astrophysics and Department of Astronomy, Yonsei University, Seoul 120-749, Korea}
\altaffiltext{2}{Space Telescope Science Institute, 3700 San Martin Drive, Baltimore, MD 21218, U.S.A.}
\altaffiltext{3}{Department of Astronomy and Space Science, ARCSEC, Sejong University, Seoul 143-747, Korea}
\altaffiltext{4}{To whom correspondence should be addressed. E-mail: ywlee2@yonsei.ac.kr}

\begin{abstract}
There is a growing body of evidence for the presence of multiple stellar populations in some globular clusters, including NGC~1851. For most of these peculiar globular clusters, however, the evidence for the multiple red giant-branches (RGBs) having different heavy elemental abundances as observed in $\omega$~Centauri is hitherto lacking, although spreads in some lighter elements are reported. It is therefore not clear whether they also share the suggested dwarf galaxy origin of $\omega$~Cen or not. Here we show from the CTIO 4m $UVI$ photometry of the globular cluster NGC~1851 that its RGB is clearly split into two in the $U - I$ color. The two distinct RGB populations are also clearly separated in the abundance of heavy elements as traced by Calcium, suggesting that the type II supernovae enrichment is also responsible, in addition to the pollutions of lighter elements by intermediate mass asymptotic giant branch stars or fast-rotating massive stars. The RGB split, however, is not shown in the $V - I$ color, as indicated by previous observations. Our stellar population models show that this and the presence of bimodal horizontal-branch distribution in NGC~1851 can be naturally reproduced if the metal-rich second generation stars are also enhanced in helium.

\end{abstract}

\keywords{globular clusters: individual (NGC~1851) --- globular clusters: formation --- stars: abundances --- stars: evolution --- stars: horizontal-branch}

\section{INTRODUCTION}
The first photometric evidence for the presence of multiple stellar populations in globular clusters came from the discoveries of the discrete distributions of stars on the red giant-branch (RGB) and main-sequence (MS) of the most luminous globular cluster $\omega$~Cen \citep{lee99,bed04}. The fact that the distribution of RGB stars is not just showing a spread but is discrete is a compelling evidence for the heavy elements enrichment and the formation of successive metal-enhanced generations of stars in proto-$\omega$~Cen. The similarity of this feature to that of the M54, suggested as a nuclear star cluster of the Sagittarius dwarf galaxy \citep{lay97}, together with other peculiarities of $\omega$~Cen, has led the community to conclude that $\omega$~Cen was once part of a more massive dwarf galaxy that merged with the Milky way, as the Sagittarius dwarf galaxy is in the process of doing now \citep{lee99,bek03}. In recent years, more and more evidence is reported for the presence of double or multiple stellar populations in other globular clusters, such as NGC~2808 \citep{pio07}, NGC~1851 \citep{mil08}, NGC~6388 \citep{mor09}, and M22 \citep{mar09,dac09}. For most of these peculiar globular clusters, however, the evidence for the discrete distribution of heavy elements as observed in the RGB of $\omega$~Cen is hitherto lacking, although spreads in some lighter elements \citep[][and references therein]{car08} and helium \citep{lee05,dan05,pio07,yoon08} are reported. Therefore, the case of $\omega$~Cen, and now that of M22 \citep{mar09,dac09},  are still viewed as exceptional, and the presence of chemical inhomogeneity in other globular clusters is largely considered as due to the pollution mechanisms expected in normal globular clusters, such as the winds from the intermediate mass asymptotic giant branch (AGB) stars or fast-rotating massive stars \citep{ven08,dec07}.

The purpose of this Letter is to report that one of these globular clusters, NGC~1851, is showing a clear split in the RGB from our CTIO 4m $UVI$ photometry. This is compared with the $Ca$-$by$ photometry \citep{jleea09} to confirm that the two distinct RGB populations are different in the abundance of heavy elements. Stellar population models are then constructed to show that, when the metal-rich second generation stars are also enhanced in helium abundance, the split of the RGB discovered in $U - I$ color would not be detected in usual optical colors, such as $V - I$ used in the $HST/ACS$ survey. 

\section{OBSERVATIONS AND COLOR-MAGNITUDE DIAGRAMS}
Our observations were performed using the CTIO 4m Blanco telescope during 2007 November 13$-$16 and 2008 October 27$-$31. The telescope was equipped with the Mosaic II CCD Imager consisting of eight 2k $\times$ 4k SITe CCDs providing a plate scale of 0.27 arcsec pixel$^{-1}$ and a field of view of 36 $\times$ 36 arcmin$^2$. All of our science frames were obtained under photometric conditions. The total exposure times for $UVI$ were 3990, 676, and 573s, respectively, split into short, intermediate, and long exposures in each band. NGC~1851 was placed on chip 6, approximately 4.5 arcmin South and 6 arcmin East from the CCD center. Several standard fields \citep{lan92,lan07} were also observed during our observing runs. The IRAF\footnote{IRAF is distributed by the National Optical Astronomy Observatory, which is operated by the Association of Universities for Research in Astronomy, Inc., under cooperative agreement with the National Science Foundation.} MSCRED package was used for preprocessing including bias correction and flat fielding. PSF photometry was then carried out using DAOPHOT II/ALLSTAR\citep{ste87}, and DAOGROW was used for aperture corrections \citep{ste90}.

Figure 1 shows color-magnitude diagrams (CMDs) of NGC~1851 in ($U$, $U - I$) and ($V$, $V - I$) planes. Stars within 3.6 arcsec and outside of 4.5 arcmin from the cluster center are excluded from the CMDs to reduce blending effects and the field star contaminations, respectively. To examine the CMD features more carefully, we adopted the ``separation index'' \citep{ste03} for selecting stars that are relatively less affected by adjacent starlights. All the stars in our CMDs lie well within chip 6, and therefore our CMDs are not subject to any uncertainty stemming from the possible chip to chip variations of the mosaic CCDs. Open circles denote RR Lyrae variables in our program field among those identified by \citet{wal98}, plotted at random phase of pulsation. The most remarkable feature of Figure 1 is the presence of two distinct RGBs in the ($U$, $U - I$) CMD. The discrete distribution is clear from the sub giant-branch (SGB) to the tip of RGB where the mean separation on the RGB is $\sim$0.27 mag in $U - I$. When measured at given I magnitude, this value is reduced to $\sim$0.20 mag. For the bright RGB stars, some early hints for this feature were noted by \citet{cala07} from the Str\"omgren ($m1$, $u - y$) CMD, and by \citet{jleea09} from the $Ca$-$by$ photometry (e.g. see their Figure 1). Note that \citet{mil08} only discovered a split in the SGB, and the split of RGB we discovered in this paper was not detected in their $HST/ACS$ photometry employing F606W and F814W passbands. Similarly, the RGB split is not apparent in our ($V$, $V - I$) CMD (Figure 1b). This is most likely because the $U$-band is more sensitive to metal abundance variation than other passbands, as more metal atomic and molecular lines are located in the $U$-band (see section 3).

Given the small foreground reddening value of E($B - V$) $=$ 0.02 \citep{har96} toward NGC~1851, it is very unlikely that the differential reddening has caused the double RGBs. The color difference between the two RGBs in $U - I$ color, at given $I$ mag of the horizontal-branch (HB) level, is $\sim$0.20 mag, which is about three times larger than the maximum color difference expected in the extreme situation where one group of stars are all reddened by E($B - V$) $=$ 0.02, while the other group has E($B - V$) $=$ 0.00. Also, if the observed color difference in $U - I$ color is due to the differential reddening, it would result in the $V - I$ color difference of 0.10 mag, which would have been detected along the RGB in our ($V$, $V - I$) CMD. Similar spatial distributions of stars on the bluer and redder RGBs also indicate that the differential reddening, if any, is not likely the cause of the double RGBs.

\section{DISCUSSION}
We have shown that the RGB of NGC~1851 is split into two distinct subpopulations. The fact that the distribution of RGB stars does not just show a spread but is discrete can naturally eliminate the possibilities such as (1) star formation from not well mixed inhomogenous interstellar matter, (2) differential reddening (see also section 2), and (3) photometric errors, as all of these would produce a $spread$ in color rather than a discrete distribution. Consequently, the most plausible interpretation for the double RGBs is that the NGC 1851 underwent metal enrichment in its early stage of evolution and successively formed metal-enhanced second generation stars. Spectroscopic observations show star-to-star abundance variations of the lighter elements (elements lighter than Si, such as N, O, Na, \& Al) in NGC~1851 \citep{hes82,yon08,yon09}. This is generally interpreted as a result of pollution from winds of intermediate-mass AGB stars \citep{ven08} or fast-rotating massive stars \citep{dec07}. Based on the elemental abundances observed by \citet{yon08} and \citet{yon09} for eight bright RGB stars, we can estimate the differences in the lighter elements between the two RGB populations. Analysis of these data by \citet[][see their Figure 3]{jleea09} indicates that, while N, Na, and Al are all enhanced in redder RGB ($\Delta$[N/Fe] $\approx$ 0.47, $\Delta$[Na/Fe] $\approx$ 0.53, $\Delta$[Al/Fe] $\approx$ 0.20), O is depleted ($\Delta$[O/Fe] $\approx$ $-$0.45), and no significant variation is shown in [C/Fe].

 In order to investigate the effect of these elemental variations in the ($U - I$) color, in Figure 2, we have computed the differences in fluxes between two synthetic spectra using the spectral-synthesis program SPECTRUM \citep{gra94}, one without and the other with these enhancements (and depletion for O) in lighter elements (panels a \& b). Also shown in Figure 2 are for the two additional cases where (1) CNO and Na are enhanced by 0.3 dex while heavier elements are fixed (panels c \& d), and (2) elements heavier than Al are enhanced by 0.3 dex while lighter elements are fixed (panels e \& f). These simulations demonstrate that small variations either in lighter or heavier elements could cause significant change in ($U - I$) color. The observed variations in the lighter elements alone, however, would cause relatively small line blanketing equivalent to $\Delta$($U - I$) $\simeq$ 0.079 in terms of color difference. Despite uncertainties, taken at face value, this is only $\sim$41\% of the observed color difference [$\Delta$($U - I$) $=$ 0.195 $\pm$ 0.011] at given $I$ magnitude and suggests that other effects may also be responsible for the RGB split in ($U - I$) color. Indeed, recent $Ca$-$by$ photometry \citep{jleea09} shows that besides lighter elements variations, RGB stars of NGC~1851 also show bimodal distribution in Ca, which can only be supplied by Type II supernovae \citep[SNe II;][]{tim95}. In Figure 3, we show in our ($U$, $U - I$) CMD the ``Ca-normal'' and ``Ca-strong'' stars from \citet{jleea09}. The Ca-strong stars lie well on the redder RGB sequence, whereas the Ca-normal stars are on the bluer RGB. According to \citet{jleeb09}, the difference in Ca abundance is estimated to be $\Delta$[Ca/H] $\approx$ 0.15 dex, and other heavy elements, albeit small, are similarly enhanced in redder RGB population. We conclude therefore that the redder RGB population is richer than the bluer RGB population not only in lighter elements (N, Na, \& Al) but also in heavy elements (such as Ca, Si, Ti, \& Fe). 

In order to better understand the origin of the RGB color difference in ($U - I$), and to place stronger constraints on the chemical combinations of the two subpopulations, we have constructed stellar population models based on the updated version of the Yonsei-Yale ($Y^2$) isochrones (Yi et al., in preparation) and HB evolutionary tracks (Han et al., in preparation). Readers are referred to \citet{yoon08} and references therein for the details of our model construction. Figure 4 presents our synthetic CMDs for NGC~1851 in ($U$, $U - I$) and ($V$, $V - I$) planes. Our models are constructed under two different assumptions regarding the chemical enrichment in NGC~1851. First, we assumed that the second generation population is more enhanced in metallicity, but not in helium (Figure 4a; hereafter $\Delta$Z-only model). Note that, while \citet{cas08} and \citet{sal08} suggested the difference in only CNO abundance, here we are assuming the difference in overall metallicity of 0.15 dex as discussed above. Second, in Figure 4b, we then assumed that both metal and helium abundances are enhanced (hereafter $\Delta$Z+$\Delta$Y model). The $U$ flux is very sensitively affected by CN and NH bands, and therefore by N abundance. In order to reflect this effect in the ($U$, $U - I$) CMD of the second generation population (redder RGB), we are also including the effects of the additional line absorptions from the enhancements in N and other lighter elements as discussed above, again by using SPECTRUM. More rigorous modeling should include the effects of these lighter elements enhancements in the construction of stellar evolutionary tracks, but we note that \citet{dot07} found the enhancement in N has only negligible effect in the HR diagram morphology at globular cluster ages.

Comparison of the number ratio between the two subpopulations suggests that bluer RGB, brighter SGB, and redder HB are associated with one subpopulation, which we refer to as ``Pop-1'', while the redder RGB, fainter SGB, bluer HB are associated with the other subpopulation, which we refer to as ``Pop-2''. Pop-1 comprises about 75\% of the total population, while Pop-2 takes about 25\% of the whole population\footnote{We have compared the ratio of the two subpopulations with samples selected in different manners (e.g. entire sample and a sample selected with various separation indices) and for all cases, the ratio comes out to be similar. This is also more or less consistent with the population ratio based on the SGB stars reported in \citet{mil09}.}. The $\Delta$Z-only model in Figure 4a ($\Delta$Z = 0.0004 and $\Delta$age = 0.1 Gyr) matches well with the observed CMD from the MS through the RGB in ($U$, $U - I$) CMD. Yet, the model fails to reproduce the blue HB and the narrow RGB in ($V$, $V - I$) CMD (Figure 4a, inset). This is because metal enhancement in Pop-2 moves HB to the opposite direction, i.e., makes HB morphology redder \citep[see for example][]{lee94}. At the same time, the increase in metallicity, albeit small, will create a gap in ($V - I$) color between the two RGBs which is not shown in the observed CMD \citep{mil08}. Therefore, the $\Delta$Z-only model is in conflict with the observed CMDs of NGC~1851. Figure 4b shows that the $\Delta$Y+$\Delta$Z model ($\Delta$Z $=$ 0.0004, $\Delta$age $=$ 0.1Gyr, and $\Delta$Y $=$ 0.05) is in good agreement with the observation from the MS to the HB. Note that, in our modeling, we are just employing the standard \citet{rei77} mass-loss law, and the same mass-loss parameter $\eta$ for both normal helium and enhanced helium populations. Input parameters adopted in our $\Delta$Y+$\Delta$Z model are listed in Table 1. The enhanced helium in Pop-2 easily overcomes the effect of metallicity on HB \citep[see][]{lee05}, shifting HB morphology toward blue and reproducing the observed HB bimodality. At the same time, the increase in helium abundance in Pop-2 moves RGB slightly bluer and this effect cancels out the metallicity effect in ($V$, $V - I$) CMD, making apparently single and narrow RGB. Also, the model (Figure 4b, inset) is in better agreement with the observed SGB split\citep{mil08}.  The $U$-band is much more sensitive to metal line blanketing, and therefore the increased helium has relatively small effect on $U - I$ color of RGB. One caveat to this helium enhanced scenario is that the blue HB in our model appears to be slightly brighter ($\sim$0.05 mag) than the observation in $U$-band, although this could be due to the uncertainty in bolometric correction of the $U$-band.

The apparent helium enhancement in the second generation subpopulation in NGC~1851 is reminiscent of the cases of other globular clusters with multiple populations, including $\omega$~Cen \citep{nor04,lee05,pio05}, NGC~2808 \citep{lee05,dan05}, NGC~6388 and NGC~6441 \citep{calo07,yoon08}. Although the origin of this helium enhancement is currently not fully understood, general consensus is that it can be supplied either by SNe II \citep{nor04,pio05} or by the intermediate mass AGB stars or fast-rotating massive stars \citep{ven09,dec07}. In the case of NGC~1851, given the enhancements both in the lighter (such as N) and heavy (such as Ca) elements in the second generation population (Pop-2), all of the mechanisms seem to be responsible for the helium enhancement. A possible scenario is that, soon after the formation of the first generation stars (Pop-1), numerous SNe II explosions enriched both metal and helium of the leftover gas in the proto-NGC~1851. Winds and ejecta from the intermediate mass AGB stars may have added more helium and simultaneously enhance lighter elements. The second generation stars (Pop-2) would have then formed from the gas enriched in overall metallicity, helium, and lighter elements. We note that the present mass \citep[$\sim$$10^6$$M_\odot$;][]{pry93} of NGC~1851 is too small to retain the ejecta from numerous SN explosions \citep{bau08,dop86}. Therefore, as in $\omega$~Cen \citep{lee99,bek03}, we suggest that NGC~1851 is most likely the remaining core of more massive primeval dwarf galaxies that merged and disrupted to form the proto-Galaxy, as recently proposed by \citet{lee07} for the origin of globular clusters with the extended HB. Our result presented here is calling a new survey of globular clusters employing $UV$ and/or Calcium filters in order to detect small difference in metallicity expected in the yet to be identified globular clusters with multiple populations. High resolution spectroscopy of stars in the two distinct groups are also needed to confirm the small difference in the abundance of heavy elements.

\acknowledgments{We thank the referee Luigi Bedin for a number of helpful suggestions. Support for this work was provided by the Creative Research Initiatives Program of the Korean Ministry of Education, Science \& Technology and KOSEF, for which we are grateful. S.-J. Y. acknowledges support from KOSEF/MEST grant (2009-0080851). J.-W. L. acknowleges support from KOSEF to ARCSEC. This material is based upon work supported by AURA through the NSF under AURA Cooperative Agreement AST 0132798, as amended.}

\clearpage
\begin{figure}
\epsscale{0.7}
\plotone{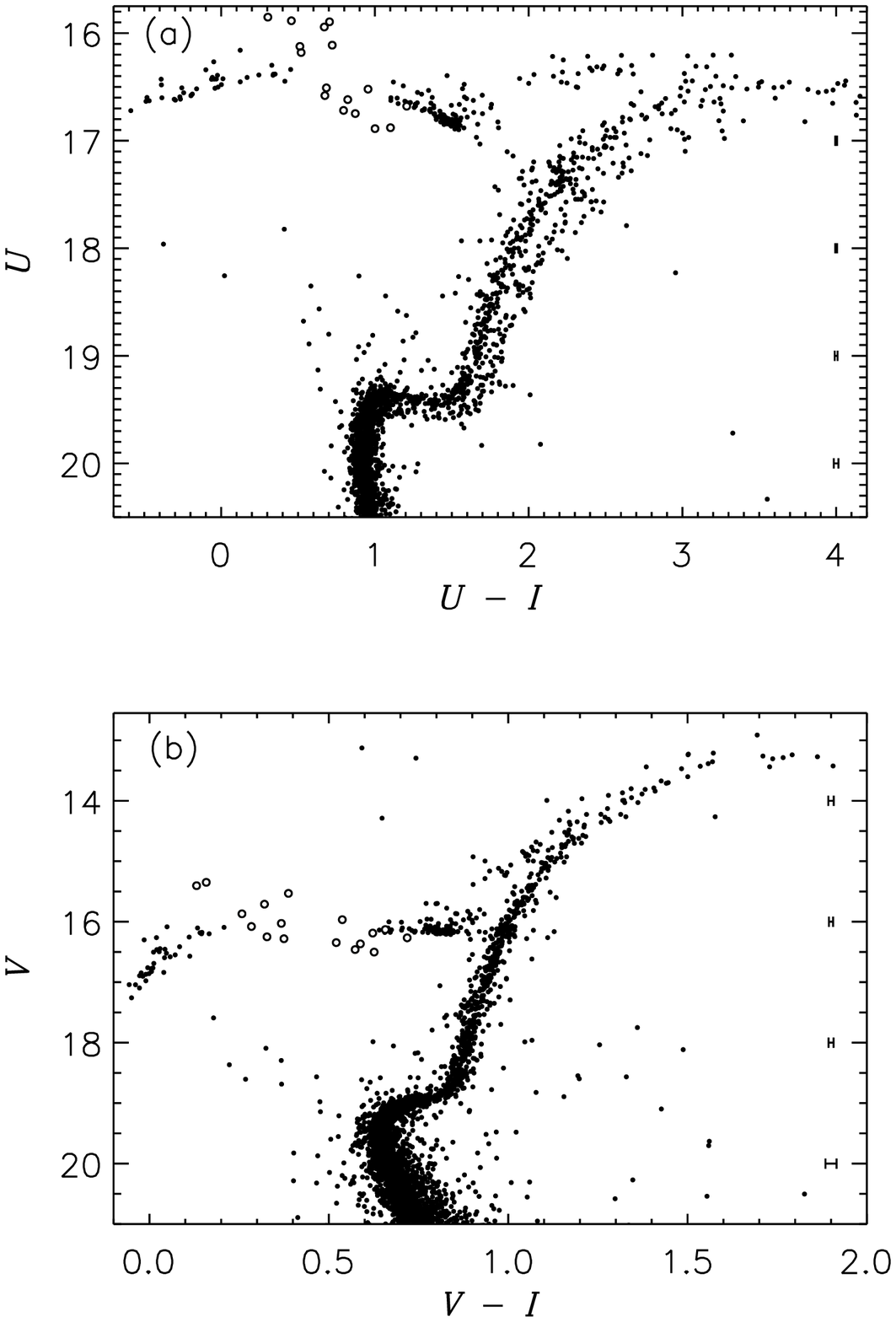}
\caption{Color-magnitude diagrams for NGC~1851. In both panels, only the stars within an annulus between 3.6 arcsec and 4.5 arcmin from the cluster center and with sep $>$ 1.0 have been plotted. Note the discrete double RGBs in the $U$ vs. $U - I$ CMD. Open circles denote RR Lyrae stars, and the photometric errors are shown. \label{fig1}}
\end{figure}

\clearpage
\begin{figure}
\epsscale{0.7}
\plotone{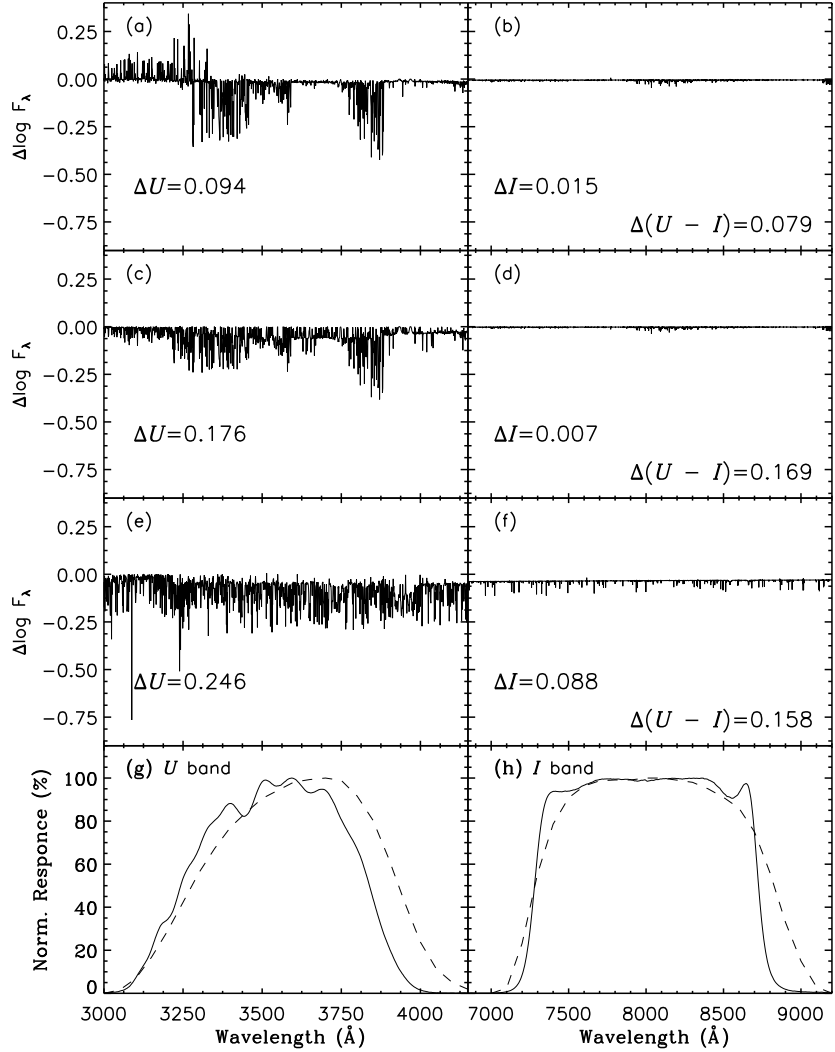}
\caption{The differences in fluxes between two synthetic spectra of RGB stars at the HB level. (a, b) For the observed variations in lighter elements ($\Delta$[N/Fe] $\approx$ 0.47, $\Delta$[Na/Fe] $\approx$ 0.53, $\Delta$[Al/Fe] $\approx$ 0.20, and $\Delta$[O/Fe] $\approx$ $-$0.45). (c, d) CNO and Na are enhanced by 0.3 dex, while heavier elements are fixed. (e, f) Elements heavier than Al are enhanced by 0.3 dex, while lighter elements are fixed. Panels (g) \& (h) are normalized responses for CTIO-4m $U$ and $I$ filters, while the dashed lines are for the Johnson filters (see text).\label{fig2}}
\end{figure}

\clearpage
\begin{figure}
\epsscale{0.7}
\plotone{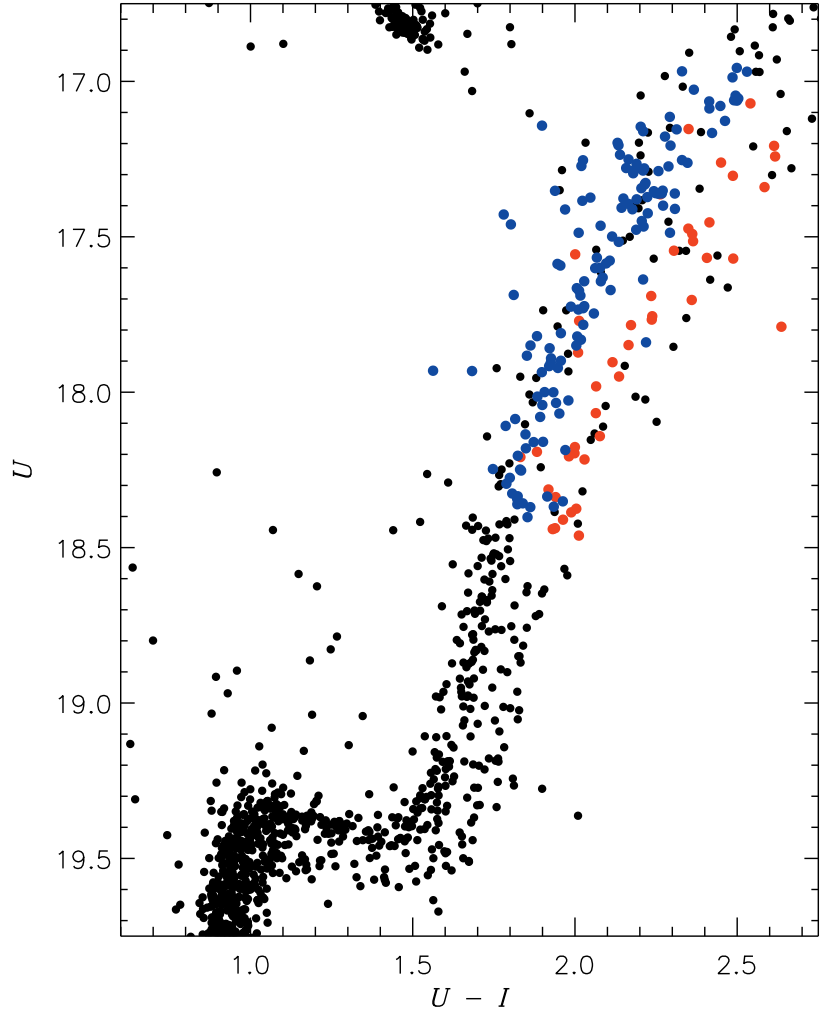}
\caption{Same as Figure 1, but zoomed around the SGB and RGB regions in $U$ vs. $U - I$ CMD. Denoted by blue and red dots are, respectively, ``Calcium-normal'' and ``Calcium-rich'' stars from \citet{jleea09}. Note that two discrete RGB sequences are well separated in calcium abundance. \label{fig3}}
\end{figure}

\clearpage
\begin{figure}
\epsscale{0.5}
\plotone{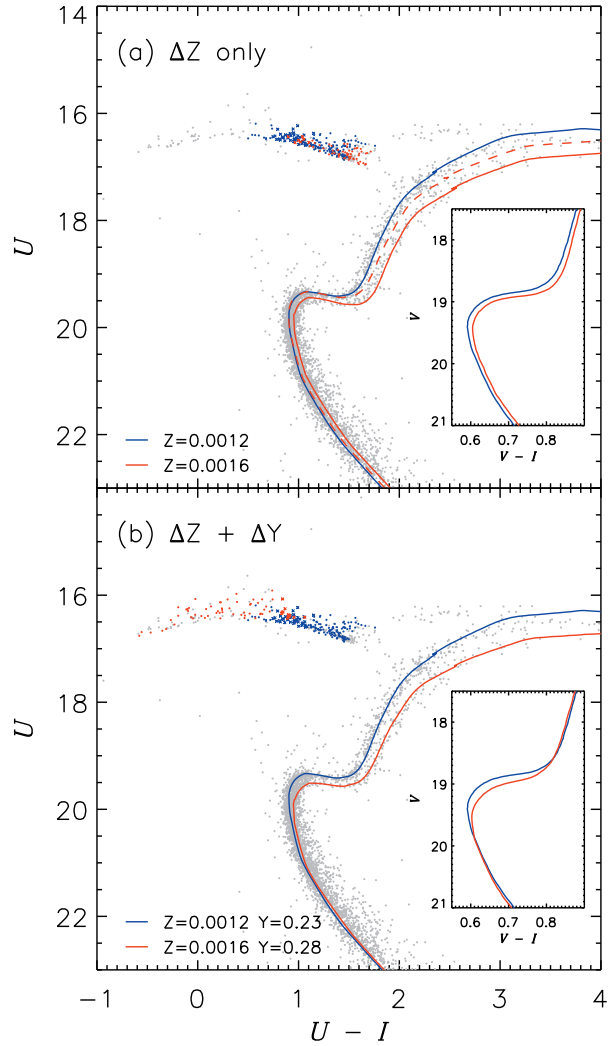}
\caption{Our population models for NGC~1851 on the observed CMD in $U$ vs. $U - I$. The models in insets are for $V$ vs. $V - I$, which should be compared with \citet{mil08} for the observed SGB split in $V$ vs. $V - I$ CMD.  Panel (a) is for the case in which only metallicity is different ($\Delta$Z $=$ 0.0004) between the two subpopulations, while panel (b) is for the case where both helium and metallicity are enhanced in Pop-2 as in Table 1. The red lines \& dots are for more metal-rich Pop-2, while the blue lines \& dots are for Pop-1. The dashed line in panel (a) shows the line blanketing effect of lighter elements only (see text). The crosses denote RR Lyrae stars. The apparent distance modulus of (m $-$ M)$_U$ $=$ 15.61 is adopted, and the color-T$_{\rm eff}$ transformation and bolometric correction from \citet{gre87} was used.\label{fig4}}
\end{figure}

\clearpage
\begin{table}
\caption{Input parameters adopted in our best simulation of NGC~1851}
\begin{center}
\begin{tabular}{cccccc}
\tableline
Parameter & Population 1 & Population 2\\
\tableline
$Z$ & 0.0012 & 0.0016 \\
$Y$ & 0.2324 & 0.282 \\
$[$$\alpha$/Fe$]$  & 0.3 & 0.3 \\
Age & 10.7 Gyr & 10.6 Gyr \\
$\eta$\tablenotemark{a} & 0.59 & 0.59 \\
$\Delta$$M$\tablenotemark{b} & 0.2337 & 0.2218 \\
$\sigma_{M}$\tablenotemark{c} & 0.015 & 0.015 \\
Population ratio & 0.7 & 0.3 \\
\tableline
\end{tabular}
\end{center}
\tablenotetext{a}{\citet{rei77} mass-loss parameter}
\tablenotetext{b}{Mean mass-loss on the RGB ($M_\odot$)}
\tablenotetext{c}{Mass dispersion on the HB ($M_\odot$)}
\end{table}

\end{document}